\documentclass[12pt]{article}
\usepackage[english,german,french,polish]{babel}
\usepackage[T1]{fontenc}
\usepackage{amsfonts}

\begin{document}

\selectlanguage{english}

\baselineskip 0.76cm
\topmargin -0.6in
\oddsidemargin -0.1in

\let\ni=\noindent

\renewcommand{\thefootnote}{\fnsymbol{footnote}}

\newcommand{\SM}{Standard Model }

\pagestyle {plain}

\setcounter{page}{1}

\pagestyle{empty}

~~~

\begin{flushright}
IFT-- 06/24
\end{flushright}

\vspace{0.5cm}

{\large\centerline{\bf A satisfactory empirical mass sum rule}}

{\large\centerline{\bf for charged leptons{\footnote{Work supported in part by the Polish Ministry of  Science and Education, grant 1 PO3B 099 29 (2005-2007). }}}} 

\vspace{0.4cm}

{\centerline {\sc Wojciech Kr\'{o}likowski}}

\vspace{0.3cm}

{\centerline {\it Institute of Theoretical Physics, Warsaw University}}

{\centerline {\it Ho\.{z}a 69,~~PL--00--681 Warszawa, ~Poland}}

\vspace{0.5cm}

{\centerline{\bf Abstract}}

\vspace{0.2cm}

In the framework of a mass formula proposed previously (transforming in a specific way three 
free parameters into three masses), a simple empirical sum rule for three charged-lepton 
masses is found, predicting $m_\tau = 1776.9926$ MeV, when the experimental values of $m_e$ 
and $m_\mu$ are used as an input. The experimental value to be compared with is $m_\tau = 
1776.99^{+0.29}_{-0.26}$ MeV. This satisfactory sum rule (equivalent to a simple parameter 
constraint in the mass formula) is linear in masses and involves integers as its coefficients. The 
author believes that such a simple and precise mass sum rule for charged leptons may help in the process of developing realistic models for mass spectra of fundamental 
fermions. In the second part of the paper, another equivalent parametrization of the charged-lepton mass formula is described, corresponding to an oscillatory picture of their mass matrix, where two 
matrices appear playing the role of annihilation and creation operators in the generation space. 
 
\vspace{0.5cm}

\ni PACS numbers: 12.15.Ff , 12.90.+b  .

\vspace{0.6cm}

\ni November 2006  

\vfill\eject

~~~
\pagestyle {plain}

\setcounter{page}{1}

\vspace{0.4cm}

\ni {\bf 1. Introduction}

\vspace{0.4cm}

Some time ago we have proposed for charged leptons $e, \mu, \tau $ the following empirical mass formula [1]:

%rownanie 1*
\begin{equation}
m_{l_i}  =  \mu\, \rho_i \left(N^2_i + \frac{\varepsilon -1}{N^2_i} - \xi\right) \;\;(l_i = e, \mu,\tau \;,\;i = 1,2,3)
\end{equation}

\ni or explicitly:

\vspace{-0.2cm}

%rownanie 2*
\begin{eqnarray}
m_e & = & \frac{\mu}{29} (\varepsilon - \xi) \,, \nonumber \\
m_\mu & = & \frac{\mu}{29}\,4\left( \frac{80 +\varepsilon}{9} - \,\xi \right) \,, \nonumber \\
m_\tau & = & \frac{\mu}{29}\, 24\left( \frac{624 + \varepsilon}{25} - \,\xi \right) \,, 
\end{eqnarray}

\vspace{-0.2cm}

\ni where 

\vspace{-0.2cm}

%rownanie 3*
\begin{equation}
N_i = 1 \;,\; 3 \;,\; 5 \;\;\;,\;\;\; \rho_i = \frac{1}{29} \;,\; \frac{4}{29} \;,\; \frac{24}{29} \;\;\;(i = 1,2,3)
\end{equation}

\ni ($\sum_i \rho_i = 1$). The fractions $\rho_i$ have been called {\it generation-weighting factors}. Here,  $ \mu, \varepsilon $ and $\xi $ are free parameters determined precisely from the charged-lepton experimental masses $m_e, m_\mu $ and $m_\tau $ [2] or {\it vice versa}. Thus, the mass formula (1) or (2) is a specific transformation of three variables into three variables, giving no independent mass predictions, {\it unless} the free parameters are constrained. To fit the free parameters, we can use the following equations inverse to the mass formulae (2):

\vspace{-0.2cm}

%rownanie 4*
\begin{eqnarray}
\mu & = & 29\, \frac{25}{9216} \left[m_\tau - \frac{6}{25}(27 m_\mu - 8m_e)\right] = 86.0076\;{\rm MeV} \,, \nonumber \\
\varepsilon & = & 10\, \frac{m_\tau - \frac{6}{125}(351 m_\mu - 904 m_e)}{m_\tau - \frac{6}{25}(27 m_\mu - 8m_e)} = 0.174069 \,, \nonumber \\
\xi & = & 10\, \frac{m_\tau - \frac{6}{125}(351 m_\mu - 136 m_e)}{m_\tau - \frac{6}{25}(27 m_\mu - 8m_e)}   = 0.0017706 \,, 
\end{eqnarray}

\ni where the experimental masses [2] are applied (for $m_\tau $ its experimental central value 1776.99 MeV is used). Note that due to Eqs. (4) the parameters $\mu, \mu\,\varepsilon $ and $\mu\,\xi $ are linear in masses.

In the approximation, where the small parameter $\xi $ is put exactly zero {\it i.e.}, the parameter constraint $\xi = 0$ is imposed, we get from the third Eq. (4) the following charged-lepton mass sum rule:

%rownanie 5*
\begin{equation}
m_\tau = \frac{6}{125} \left(351m_\mu - 136 m_e\right) = 1776.80\;{\rm MeV} \,.
\end{equation}

\ni This approximate {\it prediction} for $m_\tau $ is in a good agreement with the experimental value $m_\tau = 1776.99^{+0.29}_{-0.26}$ MeV [2]. We can also evaluate the parameters $\mu $ and $\varepsilon $ that in this approximation become

%rownanie 6*
\begin{equation}
\mu = \frac{29 (9m_\mu - 4 m_e)}{320} = 85.9924\;{\rm MeV} \;\;\;,\;\;\;
\varepsilon = \frac{320 m_e}{9 m_\mu - 4 m_e} = 0.172329\,.
\end{equation}

\ni In Eqs. (5) and (6), only the experimental values of $m_e$ and $m_\mu $ [2] are used as an input.

\vspace{0.3cm}

\ni {\bf 2. Satisfactory mass sum rule}

\vspace{0.3cm}

It is exciting enough to observe through a direct inspection of the approximate mass sum rule (5) that we can easily formulate a {\it simple} charged-lepton mass sum rule {\it predicting perfectly} the central value $1776.99$ MeV of experimental mass $m_\tau $. In fact, replacing  in Eq. (5) the coefficient 136 at the small $m_e$ by 128, we obtain

%rownanie 7*
\begin{equation} 
m_\tau = \frac{6}{125} (351m_\mu -  128 m_e) = 1776.9926\;{\rm MeV} \,,
\end{equation}

\ni where the experimental masses $m_e = 0.5109989$ MeV and $m_\mu = 105.65837$ MeV [2] are used as an input (if, alternatively, 136 were replaced by 129, then $m_\tau $ would be nearly equally excellent: 1776.9681 MeV; then, the parameter constraint (9), $\varepsilon = 97 \xi $, would be substituted by $7 \varepsilon = 775 \xi$; see Appendix for a more formal argument for 128). The prediction (7) of $m_\tau $ is even closer to 1776.99 MeV than the prediction $m_\tau = 1776.9689$ MeV [3] following (as one of two solutions) from the wonderful nonlinear Koide equation [4]. Note that Eq. (7) is linear and, like Eq. (5), involves integers as its coefficients.

It is not difficult to find out the parameter constraint equivalent to the satisfactory mass sum rule (7) {\it i.e.}, the constraint to be imposed on the parameters $\mu, \varepsilon $ and $\xi  $  (appearing in the mass formulae (2)) in order to deduce this mass sum rule. A simple calculation applying Eqs. (2) gives

%rownanie 8*
\begin{equation} 
125\,m_\tau - 6\,(351m_\mu -  128 m_e) = - 48 \frac{\mu}{29} \left(\varepsilon - 97 \xi \right)\,.
\end{equation}

\ni Thus, in the framework of transformation (2), the mass sum rule (7) is equivalent to the simple parameter constraint

%rownanie 9*
\begin{equation}
\varepsilon = 97 \xi \,.
\end{equation}

From the first and second mass formulae (2) we infer that

%rownanie 10*
\begin{equation}
\frac{\mu}{29} = \left( 1 -\frac{\xi}{10} \right)^{-1} \frac{9m_\mu - 4 m_e}{320} \;\;\;,\;\;\;
\varepsilon - \xi = \frac{29 m_e}{\mu} = \left( 1 -\frac{\xi}{10} \right)\,\frac{320 m_e}{9m_\mu - 4 m_e} \,.
\end{equation}

\ni Then, the second Eq. (10) together with Eq. (9) gives 

%rownanie 11*
\begin{equation}
\xi = \frac{10 m_e}{27 m_\mu - 11 m_e} = 0.00 17948 \,. 
\end{equation}

\ni Hence,

%rownanie 12*
\begin{equation}
\varepsilon = 97 \xi  = 0.1740927
\end{equation}

\ni and

%rownanie 13*
\begin{equation}
\mu = \frac{29  m_e}{\varepsilon - \xi} = \frac{29  m_e}{96  \xi} = 86. 007807\;{\rm MeV} \,.
\end{equation}

\ni Here, we use only the experimental values of $m_e$ and $m_\mu $ [2]  as an input. The present values of $\mu, \varepsilon $ and $\xi $, corresponding to $m_\tau = 1776.9926$ MeV, follow also from Eqs. (4) if $m_\tau = 1776.99$ MeV is replaced there by our $m_\tau = 1776.9926$ MeV.   

\vspace{0.4cm}

\ni {\bf 3. Oscillatory picture}

\vspace{0.4cm}

The charged-lepton mass formula (1) or (2) can be presented also in an equivalent form which makes use of some different parameters $A,B$ and $C$ corresponding to an oscillatory picture, where the charged-lepton mass matrix is built up from two matrices $a$ and $a^\dagger $ playing the role of annihilation and creation operators in the three-dimensional generation space of $e, \mu$ and $\tau$.

To this end, consider the following matrices acting in the three-dimensional generation space of $e, \mu, \tau$:

%rownanie 14*
\begin{equation}
a = \left( \begin{array}{ccc} 0 & 1 & 0 \\ 0 & 0 & \sqrt2 \\ 0 & 0 & 0 \end{array}\right)  \; ,\;a^\dagger = \left( \begin{array}{ccc} 0 & 0 & 0 \\ 1 & 0 & 0 \\ 0 & \sqrt2 & 0 \end{array}\right)  \;.
\end{equation}

\ni They satisfy the relations

%rownanie 15*
\begin{equation} 
n \equiv a^\dagger\, a = \left( \begin{array}{ccc} 0 & 0 & 0 \\ 0 & 1 & 0 \\ 0 & 0 & 2  \end{array}\right)  \;\;,\;\; [a\,,\, n] = a \;,\; [a^\dagger\,,\, n] = -a^\dagger \;,\; a \,a^\dagger = \left( \begin{array}{ccc}  1 & 0 & 0 \\ 0 & 2 & 0 \\0 & 0 & 0  \end{array}\right) \;,
\end{equation}

\ni and
 
%rownanie 16*
\begin{equation} 
a^2 = \left( \begin{array}{ccc} 0 & 0 & \sqrt{2} \\ 0 & 0 & 0 \\ 0 & 0 & 0  \end{array}\right)  ,\, a^{\dagger\,2} = \left( \begin{array}{ccc} 0 & 0 & 0 \\ 0 & 0 & 0 \\ \sqrt{2} & 0 & 0  \end{array}\right)  ,\, 
a^{3} = 0 ,\; a^{\dagger\,3} = 0  \,,
\end{equation}

\ni though $[ a\,,\, a^\dagger] =$ diag(1,1,-2) $ \neq $ {\bf 1}. Thus, denoting the generation triplet of 
charged-lepton fields by

%rownanie 17*
\begin{equation}
\psi (x) = \left(\begin{array}{l} \psi_e(x) \\ \psi_\mu(x) \\ \psi_\tau(x) \end{array}\right)= | 0 \!>\, \psi_e(x) + | 1 \!>\, \psi_\mu(x) + | 2 \!>\, \psi_\tau(x) \,,
\end{equation}

\ni where 

%rownanie 18*
\begin{equation}
| 0 \!>\, = \left(\begin{array}{c} 1 \\ 0 \\ 0 \end{array}\right) \;,\; 
| 1 \!>\, = \left(\begin{array}{c} 0 \\ 1 \\ 0 \end{array}\right) \;,\; 
| 2\!>\, = \left(\begin{array}{c} 0 \\ 0 \\ 1 \end{array}\right) \;,
\end{equation}

\ni we obtain from Eqs. (14) such operating relations for $a^\dagger $ and $a$ as they ought to be for creation and annihilation operators within the charged-lepton generation triplet:

%rownanie 19*
\begin{equation}
a^\dagger |0\!>\, = |1\!>\, \;,\; a^\dagger |1\!>\, = \sqrt{2}|2\!>\, \;,\; a^\dagger |2\!>\, = 0 
\end{equation}

\ni and

%rownanie 20*
\begin{equation}
a |2\!>\, = \sqrt2|1\!>\, \;,\; a |1\!>\, = |0\!>\, \;,\; a |0\!>\, = 0  \,.
\end{equation}

\ni Here, the numbers $n_i  = 0,1,2$ labelling the kets $|n_i\!>\,\;(i=1,2,3)$ are eigenvalues of the formal occupation-number operator $n \equiv a^\dagger a$ so, $n\,|n_i\!> = n_i\,|n_i\!>\,$. They may be used to label three generations of charged leptons in place of the indices $i = 1,2,3$. Note that the matrix

%rownanie 21*
\begin{equation} 
N \equiv {\bf 1} + 2 n = \left( \begin{array}{ccc}  1 & 0 & 0 \\ 0 & 3 & 0 \\ 0 & 0 & 5 \end{array}\right)  
\end{equation}

\ni has the eigenvalues $N_i = 1,3,5$.
 
Now, we will demonstrate that the charged-lepton mass formula (1) or (2) is equivalent to the conjecture that the charged-lepton mass matrix in the flavor representation,

%rownanie 22*
\begin{equation} 
M = \left( \begin{array}{ccc}  m_e & 0 & 0 \\ 0 & m_\mu & 0 \\ 0 & 0 & m_\tau \end{array}\right) \,,  
\end{equation}

\ni can be built up from the operators $a$ and $a^\dagger$ (Eqs. (14)) as follows:

%rownanie 23*
\begin{eqnarray} 
\left( \begin{array}{ccc}  m_e & 0 & 0 \\ 0 & m_\mu & 0 \\ 0 & 0 & m_\tau \end{array}\right) & = & \rho^{1/2} \left( A\, a^\dagger \,a + B\, a\, a^\dagger + C\,{\bf 1}\right) \rho^{1/2} \nonumber \\
& = &\frac{1}{29}\, \left( \begin{array}{ccc} B+C & 0 & 0  \\ 0 & 4A + 8B + 4C & 0 \\ 0 & 0 & 48A + 24C  \end{array}\right)  \;,
\end{eqnarray}

\ni where $A, B$ and $C$ are massdimensional free parameters, while

%rownanie 24*
\begin{equation} 
\rho = \left( \begin{array}{ccc}  \rho_1 & 0 & 0 \\ 0 & \rho_2 & 0 \\ 0 & 0 & \rho_3 \end{array}\right)  
\end{equation}

\ni with $\rho_i$ as given in Eq. (3) (Tr$\rho = \sum_i \rho_i = 1$). The conjecture (23) can be readily rewritten as

%rownanie 25*
\begin{eqnarray}
m_e & = & \frac{1}{29} \left(B + C\right) \,, \nonumber \\
m_\mu & = & \frac{1}{29}\,4 \left(A + 2B + C \right) \,, \nonumber \\
m_\tau & = & \frac{1}{29}\,24 \left(2A + C \right) \,. 
\end{eqnarray}

The above statement of equivalence is true, because -- similarly as the mass formulae (2) -- the new mass formulae (25) determine there three free parameters $A, B$ and $C$ in terms of three masses $m_e, m_\mu$ and $m_\tau$ or {\it vice versa} (there are no independent mass predictions, {\it unless} the new free parameters are constrained). Solving Eqs. (25) with respect to their free parameters, we get

%rownanie 26*
\begin{eqnarray}
\frac{1}{29}\,A & = & \frac{1}{72} \left(m_\tau + 6 m_\mu - 48 m_e\right) = 33.144615\;{\rm MeV} \,, \nonumber \\
\frac{1}{29}\,B & = & \frac{1}{72} \left(-m_\tau + 12 m_\mu - 24 m_e\right) = \; -7.2410213\;{\rm MeV}\,, \nonumber \\
\frac{1}{29}\,C & = & \frac{1}{72} \left(m_\tau - 12 m_\mu + 96 m_e\right) = \, 7.7520202\;{\rm MeV} \,, 
\end{eqnarray}

\vspace{0.1cm}

\ni where the experimental values of $m_e, m_\mu$ and $m_\tau$ [2] are used (for $m_\tau $ its central value 1776.99 MeV is applied).

Making use of the first Eq. (4) and the mass formulae (25), we obtain

%rownanie 27*
\begin{equation} 
\frac{9216}{25}\,\frac{\mu}{29} = m_\tau - \frac{6}{25}\,(27 m_\mu - 8 m_e) = \frac{24}{25}\,\frac{1}{29} (23A - 52B)  
\end{equation}

\ni and hence,

%rownanie 28*
\begin{equation} 
\mu = \frac{23A - 52B}{384} = 86.0076 \;{\rm MeV} \,.  
\end{equation}
 
\ni With the use of the third Eq. (4) and the mass formulae (25) we get 

%rownanie 29*
\begin{equation} 
\xi = -2\, \frac{101A + 668B + 192C}{23A - 52B}= - \frac{101A + 668B + 192C}{192 \mu} = 0.0017706 \,.  
\end{equation}

\vspace{0.1cm}
 
\ni Finally, the use of the first mass formula (2) and the first mass formula (25) gives

%rownanie 30*
\begin{equation} 
\varepsilon = \xi + \frac{29 m_e}{\mu} = \xi + \frac{B+C}{\mu} = - \frac{101A + 476B}{192 \mu} = 0.174069 \;\,. 
\end{equation}

\vspace{0.1cm}

\ni The relation (30) follows also from the second Eq. (4) and the mass formulae (25). Here, the figures for $\mu, \varepsilon $ and $\xi $, identical with those given in Eqs. (4), correspond to the experimental central value 1776.99 MeV used for $m_\tau $ (in Eqs. (28), (29) and (30), the values (26) of $A, B$ and $C$ are applied).

In the approximation, where the parameter constraint $\xi = 0$ is imposed (implying the approximate charged-lepton mass sum rule (5)), the equivalent approximate parameter constraint for $A, B$ and $C$ :

%rownanie 31*
\begin{equation} 
101 A + 668 B + 192 C = 0
\end{equation}

\ni follows from Eq. (29).

In the case of satisfactory charged-lepton mass sum rule (7), where the parameter constraint $\varepsilon = 97\xi $ is imposed, we can also find the equivalent parameter constraint for $A, B$ and $C$ by applying the mass formulae (25). Then,

%rownanie 32*
\begin{equation} 
125 m_\tau - 6 \left(351 m_\mu - 128 m_e \right) = - 24\,\frac{1}{29}(101 A + 670 B + 194 C ) \,.
\end{equation}
 
\ni Thus, in the framework of transformation (25), the mass sum rule (7) is equivalent to the parameter constraint for $A, B$ and $C$: 

%rownanie 33*
\begin{equation} 
101 A + 670 B + 194 C = 0 
\end{equation}

\ni that in turn is equivalent also to the parameter constraint (9) for $\mu, \varepsilon $ and $\xi $: $\varepsilon = 97 \xi $. The last equivalence is readily seen from Eqs. (30) and (29) giving

%rownanie 34*
\begin{equation} 
\varepsilon - 97 \xi = - \frac{1}{2\mu}(101A + 670B +194 C) \,.  
\end{equation}

\ni This relation follows also from Eqs. (8) and (31).

\vspace{0.4cm}

\ni {\bf 4. Conclusions}

\vspace{0.4cm}

In conclusion, the empirical charged-lepton mass matrix in the flavor representation has been presented in two equivalent forms:

%rownanie 35*
\begin{equation} 
\left( \begin{array}{ccc}  m_e & 0 & 0 \\ 0 & m_\mu & 0 \\ 0 & 0 & m_\tau \end{array}\right) =\mu\rho \left(N^2 + \frac{\varepsilon - 1}{N^2} - \xi \right) = \rho \left( A\, a^\dagger \,a + B\, a\, a^\dagger + C\, {\bf 1} \right)\,, 
\end{equation}

\vspace{0.1cm}

\ni where $\mu, \varepsilon, \xi$ and $A, B, C$ are free parameters, and

%rownanie 36*
\begin{eqnarray}
& & \rho = \frac{1}{29}\,\left( \begin{array}{ccc} 1 & 0 & 0 \\ 0 & 4 & 0 \\ 0 & 0 & 24 \end{array}\right)  \,,\, 
N \equiv {\bf 1} + 2n =  \left( \begin{array}{ccc} 1 & 0 & 0 \\ 0 & 3 & 0  \\ 0 & 0 & 5 \end{array}\right) \,, \nonumber \\
& & \;\;\;\; n \equiv a^\dagger a = \left( \begin{array}{ccc} 0 & 0 & 0 \\ 0 & 1 & 0 \\ 0 & 0 & 2 \end{array}\right)  \,,\,\, a\, a^\dagger = \left( \begin{array}{ccc} 1 & 0 & 0 \\  0 & 2 & 0 \\ 0 & 0 & 0 \end{array}\right)
\end{eqnarray}

\ni (Tr$\rho$ = 1), the matrices $a$ and $a^\dagger $ (Eqs. 14) playing the role of annihilation and creation operators in the three-dimensional generation space of $e, \mu, \tau$. Two forms (35) give  two equivalent mass formulae (2) and (25).The bilinear appearance of the operators $a$ and $a^\dagger $ in the mass matrix (35) may suggest the interpretation of charged-lepton differences as quantum excitations of their {\it dynamical} mass matrix.

Our crucial result here is the discovery of a simple mass sum rule (Eq. (7)) that reproduces perfectly the experimental value of $m_\tau $ through the input of experimental masses $m_e$ and $m_\mu $. In the framework of two equivalent mass formulae -- being, in fact, specific transformations between three masses and three free parameters (Eqs. (2) and (25)) -- this satisfactory mass sum rule is equivalent both to a simple constraint for $\mu, \varepsilon, \xi$ (Eq. (9)) and a little more complicated constraint for $A, B, C$ (Eq. (33)).

\vfill\eject

\baselineskip 0.74cm

{\centerline{\bf Appendix}} 

\vspace{0.25cm}

{\centerline{\it Formal argument for 128}} 

\vspace{0.25cm}

Our derivation of the satisfactory mass sum rule (7) may be described more formally as identifying in the structure of the mass formula for charged leptons a tiny combination of parameters which, if assumed to be exactly zero, gives a parameter constraint equivalent to our satisfactory mass sum rule. 

To this end, notice two relations following from Eqs. (4) that turn out to be important for charged leptons: 

$$
\frac{m_\tau -\frac{6}{125}(351 m_\mu - 128 m_e)}{m_\tau - \frac{6}{25}(27 m_\mu - 8m_e)} = -\frac{\varepsilon - 97\xi}{960} 
\eqno{\rm (A1)}
$$

\vspace{-0.1cm}

\ni and

\vspace{-0.2cm}

$$
\frac{m_\tau -\frac{6}{125}(351 m_\mu - 129 m_e)}{m_\tau - \frac{6}{25}(27 m_\mu - 8m_e)} = -\frac{7\varepsilon - 775\xi}{7680} \,. 
\eqno{\rm (A2)}
$$

\ni Their rhs's evaluated with the use of the fitted parameter values (4) are both really small in magnitude:

$$
\frac{\varepsilon - 97\xi}{960} = 2.4175\times 10^{-6} 
\eqno{\rm (A3)}
$$

\vspace{-0.1cm}

\ni and

\vspace{-0.2cm}

$$
\frac{7\varepsilon - 775\xi}{7680} = - 2.0017\times 10^{-5} \,.
\eqno{\rm (A4)}
$$

\ni Thus, also the lhs's of relations (A1) and (A2) are really small in magnitude. The magnitude smaller by one order is got by the lhs of relation (A1). Assumed to be exactly zero, it gives the satisfactory mass sum rule (7) which turns out to be equivalent to the parameter constraint (9). 

A fractional number lying between 128 and 129 for which the calculated mass $m_\tau $ is exactly equal to its (actually valid) experimental central value 1776.99 MeV is not especially interesting from our point of view restricted to integers replacing 136 at $m_e$, when the approximate mass sum rule (5) is being improved. At any rate, note that it is close to 128:

\vspace{-0.2cm}

$$
\frac{351 m_\mu -\frac{125}{6} 1776.99\;{\rm MeV}}{m_e} = 128.108\,.
\eqno{\rm (A5)}
$$

\vfill\eject
~~~~
\vspace{0.5cm}

{\centerline{\bf References}}

\vspace{0.5cm}

{\everypar={\hangindent=0.6truecm}
\parindent=0pt\frenchspacing

{\everypar={\hangindent=0.6truecm}
\parindent=0pt\frenchspacing

[1]~ W. Kr\'{o}likowski, {\it Acta Phys.~Pol.} {\bf B 37}, 2601 (2006)[{\tt hep--ph/0602018}]; and references therein; {\it cf.} also {\tt hep--ph/0604148} for an "intrinsic interpretation"\, of the previous paper, supporting the specific form (1) of mass formula.

\vspace{0.2cm}

[2]~W.M.~Yao {\it et al.} (Particle Data Group), {\it Review of Particle Physics, J.~Phys.} {\bf G 33}, 1 (2006).

\vspace{0.2cm}

[3]~W. Kr\'{o}likowski, {\tt hep--ph/0508039}.

\vspace{0.2cm}

[4]~For a recent discussion {\it cf. } Y.~Koide, {\tt hep--ph/0506247}; and references therein;  {\it cf. } also A.~Rivero and A.~Gsponer, {\tt hep--ph/0505220}.

\vfill\eject

\end{document}